\documentclass[twocolumn,showpacs,preprintnumbers,amsmath,amssymb]{revtex4}
\usepackage{graphicx}
\usepackage{dcolumn}

\def\be{\begin{equation}}
\def\ee{\end{equation}}
\def\bea{\begin{eqnarray}}
\def\eea{\end{eqnarray}}

\def\prl{Phys. Rev. Lett.}

\def\prb{Phys. Rev. B}

\begin{document}
\title{Plasticity and reversibility of structural transitions in a model solid}
\author{Arya Paul, Surajit Sengupta}
\affiliation{S.N. Bose National Centre for Basic Sciences, Block JD, 
Sector III, Salt Lake, Calcutta 700 098, India}
\author{Madan Rao}
\affiliation{Raman Research Institute, C.V. Raman Avenue, Bangalore 560 080, 
India\\
National Centre for Biological Sciences (TIFR), Bellary Road,
Bangalore 560 065, India
}
\begin{abstract}
We formulate a phenomenological elasto-plastic theory to describe a solid 
undergoing a structural transition from a square ($p4mm$) to an oblique ($p2$) 
lattice in two dimensions. Within our theory,  the components of the strain 
may be decomposed additively into separate elastic and plastic contributions. 
The plastic strain, produced when the local stress crosses a threshold, is 
governed by a phenomenological equation of motion. We investigate the 
dynamics of shape of an initially square solid as it is 
cycled through a transformation protocol consisting of 
(1) a quench across the transition (2) deformation by an external stress and 
finally (3) reverse transformation back to the parent state. We show that shape 
recovery at the end of this cycle depends on crucially on the presence 
of plasticity in components of the strain responsible for the transformation.   
\end{abstract}
\maketitle
\section{Introduction}

The question of the degree and efficiency of shape reversibility of solids 
undergoing reversible Martensitic transitions\cite{paper0} is clearly of 
considerable technological 
importance considering the diverse applications of shape memory 
or ``smart'' materials\cite{smm}. 

The necessary criterion for a material in order to exhibit shape-memory is 
that the symmetry group of the product phase forms a sub-group of the 
parent and the transformation leads to symmetry breaking\cite{kaushik}. 
In a recent publication we had explored the possible {\em sufficient} 
conditions for shape reversibility using a model system in two dimensions 
as a test case\cite{nanomart}. In our model system it was possible to 
influence the nature of the transformation by tuning the parameters of 
the interaction potential\cite{paper1}. We found that, while a 
square ($p4mm$) to a rhombic (in general, oblique - $p2$) lattice was 
reversible\cite{hatch}, the transformation to the more symmetric, triangular 
($p6$) lattice was not, in accordance with the conclusions of 
Ref.\cite{kaushik}. 
However, we found that the question of reversibility was linked to the extent 
and nature of plastic or {\em non-affine} zones (NAZs) present within the
transforming crystal. The positions of particles which belong to the NAZs 
cannot be obtained from a simple affine (shear and/or scaling) transformation
of the parent lattice. We identified two kinds of NAZs associated either 
with the non-order parameter (NOP) (e.g. volumetric) strain or with the 
order-paramter (OP) shear strain which transforms the square to a rhombic 
lattice. We argued that the $p4mm \leftrightharpoons p2$ produces only NAZs 
of the first kind while during a $p6 \to p4mm$ transition, we gave 
indirect evidence that  both kinds of NAZs are likely to be produced leading 
to irreversibility. We concluded therefore that for reversibility it is 
sufficient that NAZs associated with OP strains  not be produced. To 
support our claim, we showed that eliminating NAZs by making the solid 
stronger or the system size smaller than the typical distance between NAZs 
makes even the $p4 \to p6$ transformation reversible in 
apparent contradiction to Ref.\cite{kaushik}. 
      
In this paper, we lend further support to this claim by investigating this 
problem from the point of view of a phenomenological elasto-plastic 
theory\cite{paper2} of the square to oblique transformation. 
The elasto-plastic theory has been shown earlier to reproduce the qualitative 
features of the results of our MD simulations\cite{paper1}.
One of the advantages of this approach is that plasticity in the NOP or 
the OP sector may be introduced systematically and their effects observed 
explicitly. 

Our main results are as follows. We study the reversibility of shape of a 
solid which is first quenched from a square into the rhombic phase, deformed 
by an external stress and finally brought back to the square phase (see Fig.1).
The external control parameters are the temperature and stress. We compare 
the initial and final shapes of the solid and look for congruence. 
We show that the presence of plasticity exclusively in the NOP sector 
does not affect shape reversibility while the presence of plasticity in the 
OP sector triggers irreversibility in shape changes.

The next section is dedicated to a quick review of our elasto-plastic theory 
where we discuss the formalism and how to deduce shape at each instant of 
time from the knowledge of strains at that instant. In section III we 
present our results and we conclude in section IV indicating possible 
future directions. 

\section{Elasto-plastic theory}

In this section we formulate our elasto-plastic continuum theory for 
structural transitions in solids. Our theory allows us to determine the 
microstructure and the overall shape of a solid undergoing a square 
to a general oblique  lattice. Unlike previous work in this 
subject\cite{strn-only} we allow for the presence of non-affine,
or plastic deformations\cite{paper1,paper2}. 
 
Our main assumption in what follows is that the components of the strain 
tensor\cite{Landau} may be additively decomposed into elastic 
(affine) and plastic parts\cite{elastop}, {\it viz.} $e_{ij} = e_{ij}^A + e_{ij}^P$ where 
$i,j = x,y$. The plastic strains $e_{ij}^P$ represents the total contribution 
from, in general, space and time dependent defect fields, which, 
in the spirit of our continuum approach, need not be resolved into individual,
microscopic defects. Nevertheless, defects introduce 
multivaluedness in the particle displacement field $\mathbf u = 
(u_x(\mathbf r), u_y(\mathbf r))$ and cause the elastic parts of the strain 
to violate the St. Venant's compatibility condition\cite{baus}, 
\begin{equation}
\nabla \times (\nabla \times \Bar{\Bar e}^A)^T = 
- \nabla \times (\nabla \times \Bar{\Bar e}^P)^T \ne 0,
\label{venant}
\end{equation}
where $\Bar{\Bar e}^A$ and $\Bar{\Bar e}^P$ represents the affine 
(elastic) and the plastic strain tensors respectively. 
Note that Eq.(\ref{venant}) implies that the {\em total} strain $e_{ij}$ does 
satisfy the compatibility condition. 

The structural transition is driven by the non-linear response of the solid 
to one or more components of the strain- the order parameter (OP). For 
example, for the ($p4mm\to p2$) transition, the order parameter strains are 
$e_2 = e_{xx} - e_{yy}$ and $e_3 = e_{xy} = e_{yx}$ while the remaining 
volumetric strain $e_1 = e_{xx} + e_{yy}$ is a non-order parameter (NOP) 
strain\cite{hatch}. We shall first consider the case when plasticity 
exists only in the NOP sector. 

The transition is described by the following free energy functional: 
\begin{eqnarray}
{\cal F}  &=&  \frac{1}{2}\int\,{\rm dxdy}\,\left[ a_1(e_1^A)^2\,+\,a_2e_2^2\,+\,a_3e_3^2 - b_3e_3^4 + d_3e_3^6 \right.    \\\nonumber
          &+& \left.\,c_1(\nabla e_1^A)^2\, +\,c_2(\nabla e_2)^2\,+\,c_3(\nabla e_3)^2  \right].
\label{freen}
\end{eqnarray}
Note that in Eq.(\ref{freen}) we have made the simplifying assumption 
that the product oblique lattice is actually a rhombus so that the 
equilibrium value of $e_2 = 0$. This choice is motivated by our MD simulation 
of a particular model solid \cite {paper1} though we do not doubt that 
our theory may be easily extended to the general case. The three elastic 
constants $a_1, a_2$ and $a_3$ define the linear elasticity of the square 
(parent) phase. The connection with external control parameters such as 
temperature ($T$) is, as usual, through the temperature dependence of these 
coefficients especially $a_3 \propto (T-T^*)/T^*$ where $T^*$ is the limit 
of stability of the square solid. 
Reducing $a_3$ by cooling the solid stabilizes the rhombic 
phase (see Fig. 1).
The rest of the coefficients parametrize non-linearities and may be taken 
to be constants. 

The Lagrangian is given by\cite{strn-only}, 
\begin{eqnarray}
{\cal L} &=& \int \frac{\rho}{2}\left[(\dot u_x^2 + \dot u_y^2)\right]dxdy - {\cal F}.
\label{lag}
\end{eqnarray}
To obtain the equation of motion in the displacement fields, we need to solve 
the Euler Lagrange equation: 
\begin{equation}
\frac{d}{dt}\frac{\partial L}{\partial \dot u_i} - \frac{\partial L}{\partial u_i} = 
- \frac{\partial R}{\partial \dot u_i}.
\label{euler-lag}
\end{equation}
The Rayleigh dissipation functional \cite{Landau} ${\cal R}$ is given by 
${\cal R} = 
\frac{1}{2}\int \left[\xi (\dot e_1^A)^2 + \kappa \dot e_2^2 + \gamma \dot e_3^2\right]dxdy$, where the coefficients $\xi, \kappa$ and $\gamma$ are the 
corresponding viscosity coefficients of the system.

Using Eqs.(\ref{freen}), (\ref{lag}) and (\ref{euler-lag}), we obtain 
the following equations of motion for the affine strains $e_1^A, e_2$ and 
$e_3$:
\begin{eqnarray}
\label{e1}
\rho\ddot e_1^A &=& 
\nabla^2\left(\frac{\delta {\cal F}}{\delta e_1^A} + \frac{\delta {\cal R}}{\delta \dot e_1^A} \right) + 
\frac{\partial^2}{\partial x\partial y}\left( \frac{\delta {\cal F}}{\delta e_3} + \frac{\delta {\cal R}}{\delta \dot e_3}  \right) \nonumber\\
&+& {\cal W}^2 \left(\frac{\delta {\cal F}}{\delta e_2} + \frac{\delta {\cal R}}{\delta \dot e_2}\right),  
\label{e2} \\
\rho\ddot e_2 &=& 
\nabla^2\left( \frac{\delta {\cal F}}{\delta e_2} + \frac{\delta {\cal R}}{\delta \dot e_2}\right) + 
{\cal W}^2 \left(\frac{\delta {\cal F}}{\delta e_1^A} + \frac{\delta {\cal R}}{\delta \dot e_1^A} \right), 
\label{e3} \\
\rho\ddot e_3 &=& 
\frac{1}{4}\nabla^2\left(\frac{\delta {\cal F}}{\delta e_3} + \frac{\delta {\cal R}}{\delta \dot e_3}\right)
 + \frac{\partial^2}{\partial x\partial y}\left(\frac{\delta {\cal F}}{\delta e_1^A} + \frac{\delta {\cal R}}{\delta \dot e_1^A}\right).
\end{eqnarray}
where the ``wave'' operator ${\cal W}^2$ is defined as ${\cal W}^2 = \partial^2/\partial x^2 - \partial^2 / \partial y^2$. 
Since there is symmetry breaking in the space of the OP strains, the dynamics 
of $e_2$ and $e_3$ -- the broken symmetry variables -- is much slower than 
that of
the affine NOP strain $e_1^A$. Therefore $e_1^A$ 
reaches a steady state much faster compared to $e_2$ and $e_3$ and 
we can simplify Eqs.(\ref{e1}-\ref{e3}) by assuming that $e_1^A$ is 
slaved, at all times, to the OP strains $e_2$ and $e_3$. 
Also note that, the St. Venant's 
compatibility condition, which in two dimensions can be explicitly written as 
\begin{equation}
\nabla^2e_1 - {\cal W}^2 e_2 - 4\nabla_x\nabla_ye_3 = 0,
\label{venant-2d}
\end{equation}
may be used to eliminate $e_2$ from Eqs.(\ref{e1}-\ref{e3}). 

The final set of equations which therefore has to be solved to obtain the 
affine strains are 
\begin{eqnarray}
\label{e1-dyn}
\nabla^2 e_1^A &=& \left(\frac{4a_2-a_3}{a_1+a_2}\right)\frac{\partial^2e_3}{\partial x\partial y} - \left(\frac{a_2}{a_1+a_2}\right)\nabla^2 e_1^P. 
\label{e3-dyn} \\ 
\rho \ddot e_3 &=& \frac{1}{4}\nabla^2\left(a_3e_3 - b_3e_3^3 + d_3e_3^5 
- c_3\nabla^2e_3 \right. \\\ \nonumber
&-& \left.\ 4c_1\frac{\partial^2 e_1^A}{\partial x\partial y} 
+ \gamma\dot e_3\right) + 
\frac{\partial^2}{\partial x\partial y}\left(a_1e_1^A + \xi\dot e_1^A\right)
\end{eqnarray}

\begin{figure}
\begin{center}
\includegraphics[width=7cm]{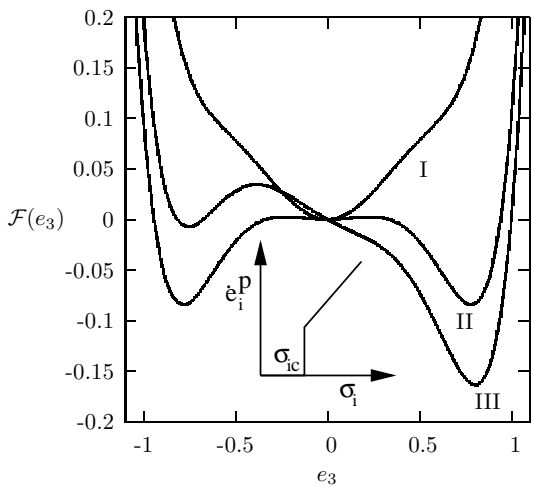}
\end{center}
\label{freplot}
\caption{
The homogeneous part of the free energy functional ${\cal F}(e_3)$ as a 
function of the OP strain $e_3$ for different values of the parameters
$a_3$ and the external stress $\sigma_3$ viz. $a_3 = 0.5, \sigma_3 = 0$ (I),
$a_3 = 0.1, \sigma_3 = 0$ (II) and $a_3 = 0.1, \sigma_3 = .1$ (III).
The rest of the parameters $b_3 = -1$ and $d_3 = 1$ throughout.
Inset shows a schematic representation of the typical dynamics of plastic 
strain $\dot e_i^P$ as a function of the local stress $\sigma_i$ used in our 
calculations. Similar dynamics is assumed for volumetric (NOP, $i = 1$) as well 
as shear (OP, $i = 3$) plasticity. The threshold stress $\sigma_{ic}$ is, of 
course, different in the two cases.}  
\end{figure}

Lastly, the dynamics of the plastic strain  $e_1^P$ needs to be specified. We 
assume that this is given by the following simple phenomenological equation of 
motion used by us (Fig. 1 inset) is, 
\begin{eqnarray} 
\label{micplas}
\dot {e}_{1}^P & = & \frac{1}{h_1}\,\,\,\sigma_{1} - c_p\nabla^2e_1^P
                      \,\,\,\,\,\,\,\,\,\,\,\,\,\,\,\,\,\,\,{\rm if}\, \vert \sigma_{1}\vert  > \sigma_{1c} \\ \nonumber
               & = & -c_p\nabla^2e_1^P \,\,\,\,\,\,\,\,\,\,\,\,\,\,\,\,\,\,\,\,\,\,\,\,\,\,\,\,\,\,\,\,\,\,\,\,\,\,\,\,\,\,{\rm otherwise} 
\end{eqnarray}
where the local volumetric stress $\sigma_1 = \delta  {\cal F}/\delta e_1^A$, 
and we have chosen a Newtonian ansatz with a simple threshold 
criterion with yield stress $\sigma_{1c}$ for simplicity. Note that 
Eq.\ref{micplas} also includes {\em strain diffusion}.  

The continuum elasto-plastic theory described above has been 
successfully used in the past\cite{paper2} to describe microstructure selection
following a quench from the square to the oblique phase. For large 
values of $h_1$ in Eq.(\ref{micplas}) $e_1^P$ is produced only at 
the parent-product interface within localized and transient NAZs. Once 
coarse-grained over distances large compared to the typical size of 
these NAZs, the twinned microstructure obtained in this limit become
identical to that seen in ``strain-only'' approaches\cite{strn-only} which do
not 
involve plasticity. On the other hand, for small values of $h_1$, 
$e_1^P$ proliferates the entire solid and effectively {\em screen}
non-local elastic interactions, completely changing the microstructure
from twinned martensite to un-twinned ferrite. In this limit the coarse-
graining length grows to become of the order of the system size
and such microstructures cannot be described within the earlier 
strain only theory\cite{strn-only}. These results closely mimic those observed 
in our MD 
simulations of the square $\leftrightharpoons$ oblique transformation
in a model solid\cite{paper1}.

In this paper we wish to study shape recovery following a 
quench- deform- reheat transformation cycle for a solid undergoing 
a structural transition in the presence of plasticity. We therefore
need to include the effect of external stress by using a modified 
Lagrangian, namely,  
\begin{equation}
\label{mod-lag}
{\cal L} = \int \frac{\rho}{2}\left[(\dot u_x^2 + \dot u_y^2)\right]dxdy - {\cal F} + f_xu_x + f_yu_y
\end{equation} 
where $f_x$ and $f_y$ are the components of the tangential forces along the 
$x$ and $y$ direction respectively applied at 
the edges of the system. The dynamical equations for the affine strains, 
obtained from this Lagrangian are
\begin{eqnarray}
\label{e3-mod}
\rho \ddot e_3 &=& \frac{1}{4}\nabla^2\left(a_3e_3 - b_3e_3^3 + d_3e_3^5 
- c_3\nabla^2e_3 \right. \\\ \nonumber
&-& \left.\ 4c_1\frac{\partial^2 e_1^A}{\partial x\partial y} 
+ \gamma\dot e_3\right) + 
\frac{\partial^2}{\partial x\partial y}\left(a_1e_1^A + \xi\dot e_1^A\right) \\
\nonumber
&+& \sigma_3 
\end{eqnarray}
Note that, the external shear stress $\sigma_3 = \frac{\partial f_x}{\partial y} + \frac{\partial f_y}{\partial x}$
does not affect the steady state equation for the 
affine NOP strain $e_1^A$. The phenomenological equation for the plastic 
strain also remains same as well. 

If, in addition, plasticity is associated with the OP strain, $\sigma_3$ 
may also give rise to plastic flow in the solid. In this work, 
we consider this possibility as well. In order to incorporate plasticity in the 
OP sector, we proceed in an identical manner. We write $e_3 = e_3^A + 
e_3^P$. We replace $e_3$ with $e_3^A$ in Eq.(\ref{e1-dyn}) and 
Eq.(\ref{e3-mod}). the dynamics of $e_3^P$ is assumed to be given by a 
phenomenological equation similar to eqn(\ref{micplas}). Finally, as before, 
the total strains $e_1$ and $e_3$ satisfy the St. Venant's compatibility 
condition Eq.(\ref{venant-2d}). 

Having specified the dynamics of the strains, it is necessary to be able to 
compute the shape of the solid from the 
strain fields in order to investigate shape reversibility. For this, we 
make use of the Kirchoff-Cessaro-Voltera \cite{baus} relation
\begin{equation}
\mathbf u(\mathbf r) = \int_{C(\mathbf r_0,\mathbf r)}[\mathbf E(\mathbf l) + (\mathbf l - \mathbf r)\times \mathbf \nabla_{\mathbf l}\times \mathbf E(\mathbf l)]\cdot d\mathbf l
\label{KCV}
\end{equation}
where $E(\mathbf l)$ is the strain tensor at position $\mathbf l$. The line 
integral is along any arbitrary contour $C(\mathbf r_0,\mathbf r)$ from a 
fixed point 
of the deformation $\mathbf r_0$ to the point of interest $\mathbf r$. The 
displacements, so evaluated, are valid upto a global translation and global 
rotation, that can be viewed as integration constants. In two-dimensions, 
Eq.(\ref{KCV}) reduces to,
\begin{eqnarray}
\label{ux}
u_x(\mathbf r) &=& \int_{\mathbf r_0}^{\mathbf r}\left[e_3 + 
(y-\xi_y)\frac{de_3}{d\xi_y} \right. \\ \nonumber 
&-& \left. (y-\xi_y)\frac{d(e_1-e_2)}{d\xi_x}\right]d\xi_y \\  \nonumber
&+& \int_{\mathbf r_0}^{\mathbf r}\left[e_1 + e_2 + 
(y-\xi_y)\frac{d(e_1+e_2)}{d\xi_y} \right. \\ \nonumber 
&-& \left.
(y-\xi_y)\frac{de_3}{d\xi_x}\right]d\xi_x 
\label{uy} \\
u_y(\mathbf r) &=& \int_{\mathbf r_0}^{\mathbf r}\left[e_1-e_2 + 
(x-\xi_x)\frac{d(e_1-e_2)}{d\xi_x} \right. \\ \nonumber 
&-& \left. (x-\xi_x)\frac{de_3}{d\xi_y}\right]d\xi_y \\  \nonumber
&+& \int_{\mathbf r_0}^{\mathbf r}\left[e_3 + (x-\xi_x)\frac{de_3}{d\xi_x} - 
(x-\xi_x)\frac{d(e_1+e_2)}{d\xi_y}\right)d\xi_x
\end{eqnarray}
Evaluating the displacements $u_x$ and $u_y$ at the boundaries will generate 
the shape of the system subjected to a global rotation and a global 
translation. Note that total strains, which satisfy the compatibility 
constraint Eq.(\ref{venant-2d}), are used to evaluate the displacements. 

\begin{figure}
\begin{center}
\includegraphics[width=8cm]{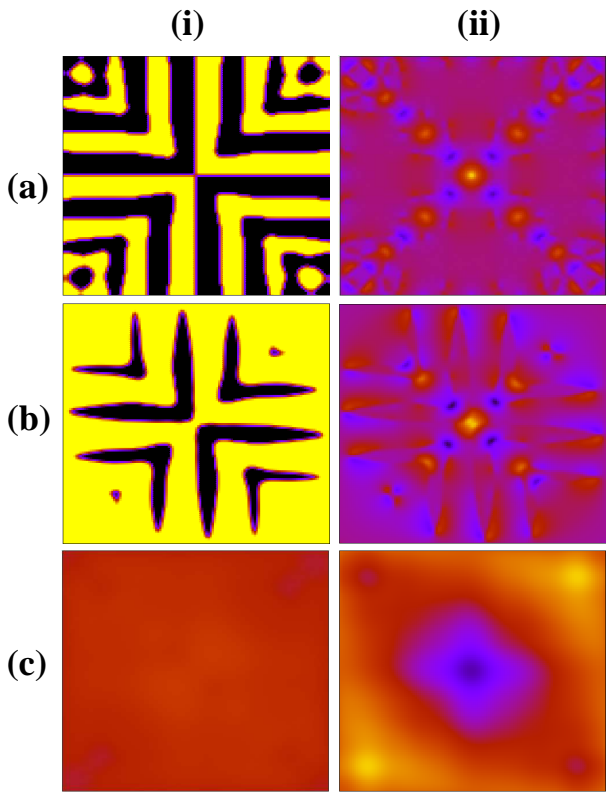}
\end{center}
\caption{ (a)(i) Plot of $e_3$ showing the quenched microstructure. The 
parameters are $\rho = 1, a_3=0.04, b_3 = -6, d_3 = 4, c_3=4,a_1 = 4, 
c_1 = 0.04, a_2 = 0.04, c_p = 0.04, \gamma = 20, \xi = 0.4, \sigma_{1c} = 1$ 
and $h_1 = 10$. (ii) Corresponding plot of the affine strain $e_1^A$. 
(b)(i) The micro- structure and (ii) $e_1^A$ when the system is placed 
under an external shear stress. The parameter $\sigma_3 = 0.5$. 
(c)(i) The order parameter when the system is ``heated'' by increasing $a_3$
 from $0.04$ to $4$. The entire system goes into a square phase phase with 
$e_3=0$. (ii) The remnant volumetric strain $e_1^A$ averages to zero. 
The color range is from -1 (black) to 0 (brown)
to 1 (yellow) for (i)(a),(b)\&(c). For (ii)(a),(b)\&(c) the corresponding
range is from -.1 (black) to .1 (yellow).  
}
\label{nop-plast}
\end{figure}

\section{Results} 
As mentioned before,
we look for shape reversibility at the conclusion of the following 
transformation protocol. 
Firstly, a finite size solid initially in the shape of a square with a parent 
(square or $p4mm$) crystal structure (Fig. 1 curve I)is quenched to get a 
rhombic martensitic product with a $p2$ structure (Fig. 1 curve II). Next, the 
product phase is  deformed (Fig. 1 curve III) with the help of an external 
shear strain $\sigma_3$. Finally, the deformed system is 
transformed back to the parent square crystal structure (Fig. 1 curve I). All 
throughout we keep track of the shape using Eqs.(\ref{ux}) and (\ref{uy}). If 
the system returns to the square shape after heating, we conclude 
that the shape is reversible, otherwise not. We have performed this 
transformation protocol for two cases: Case I- when there is only NOP 
plasticity and no OP plasticity Case II- when plasticity is present in both 
NOP and OP sector. We give below these results for the two cases, one 
after the other. 

\subsection{Case-I }
We first consider the case where plasticity exists only in the NOP sector. 
We need to solve Eqs.(\ref{e1-dyn}) and (\ref{e3-mod}) 
to get the desired quenched microstructure from the parent square phase. 
Our initial values of $e_3(\mathbf r)$ and $e_1^A(\mathbf r)$ are zero 
everywhere. For the plastic strain $e_1^P$, we assume $e_1^P(\mathbf r,0) = 
\delta(x-L/2)(y-L/2)$ at $t=0$. When quenched below the transformation 
temperature, the OP strain $e_3$ shows the formation of twins, originating 
from the center of the simulation box where the initial value of the  plastic 
strain was nonzero 
[see Fig(2)(a)(i)]. Fig(2)(a)(ii) shows the affine NOP strain $e_1^A$. The 
calculated shape of the quenched solid is shown in Fig(3)(a)(i). Although 
alternating twins are present, the overall shape of the solid continues, on 
an average, to be a square.  

We then load the microstructure using an external shear stress so that only 
one of the product variants is favored. The system prefers to go to a single 
variant in order to lower the free energy.
We stop the evolution after the system has undergone sufficient 
deformation. The plots of the OP strain $e_3$ and the affine NOP strain 
$e_1^A$ is shown in Fig(2)(b).  The shape of this deformed system is shown in 
Fig(3)(b)(i). 

Having deformed the system, we transform it back to the parent phase by 
increasing $a_3$ (temprature). The plots of the OP strain $e_3$ and the 
affine NOP strain 
$e_1^A$ is shown in Fig(2)(c). The whole system relaxes to zero OP strain. 
The final shape of the system is plotted in Fig(3)(c)(i). The system 
goes back to the parent phase, i.e., a square phase, and, at the same time
recovers its shape at the end of the complete transformation cycle. 
The plasticity in the NOP sector does not affect shape recovery because
the $e_1^P$ alternate in sign and average to zero over the entire sample. 
The shape is determined mainly by the OP strain which is 
completely reversible. 
 
\begin{figure}
\begin{center}
\includegraphics[width=6cm]{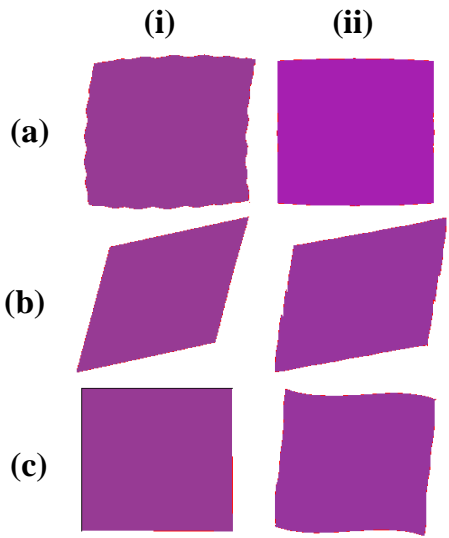}
\end{center}
\caption{(a)(i) - (c)(i) The shape of the system corresponding to quenched, 
loaded and heated microstructure in Fig. (2) and (a)(ii) - (c)(ii) Fig. (4). 
}
\label{shape}
\end{figure}
\begin{figure}
\begin{center}
\includegraphics[width=8cm]{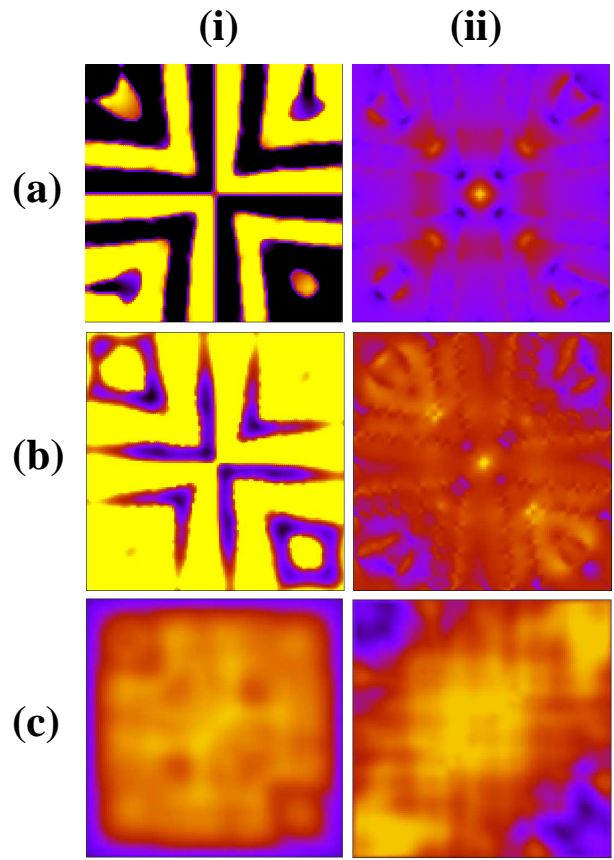}
\end{center}
\caption{(a)(i) Plot of $e_3^A$ for the quenched microstructure.(ii) 
Corresponding plot for $e_1^A$. (b)(i) Plot for $e_3^A$ for the deformed 
microstructure.(ii) Corresponding plot for $e_1^A$. (b)(i) Plot for $e_3^A$ 
after heating the system.(ii) Corresponding plot for $e_1^A$. $\sigma_{3c} 
= 2$ and $h_3 = 1$. All other parameters are same as that of Fig(2). For 
(i)(a)-(c), the color range in from -1.5 (black) to 1.5 (yellow) 
For (ii)(a)-(c), the color range goes from -0.1(black) to 0.1(yellow).
}
\label{op-plast}
\end{figure}

\subsection{Case II}
We now look into a system undergoing square to rhombic transition including  
plasticity in the OP sector as well. As before, we start with a 
solid in the shape of a square and in the $p4mm$ phase 
with a delta function NOP plasticity at the center 
of the simulation box and let the system evolve
without any external stress to obtain the microstructures. The plots 
for the affine OP strain $e_3^A$ 
and the affine NOP $e_1^P$ are shown in Fig. 4(a). Note that these, 
as well as the calculated external shape, shown in Fig. 3. (a)(ii), are not 
significantly different from those obtained in Case I.

The quenched microstructure, so obtained, is then, as before, deformed with an 
external shear stress as in the previous case. Unlike Case I, however, the 
external deformation now has two effects. Firstly, as before, the external 
stress favors one of the twin variants over the other and therefore alters 
the microstructure by changing the distribution of the variants. Secondly,
the external stress may cause the solid to flow, so that a part of the 
deformation, would now be plastic. 
The amount of strain which appears as the plastic strain in the OP sector, 
of course, depends on the parameters eg $\sigma_{3c}$ - the threshold stress 
and $h_3$ - the rate of plastic strain production. In the present work, we 
have used a low value of the stress threshold to over-emphasize the plastic 
contribution for the sake of illustration. In reality, the contribution of 
the plastic 
strain may be smaller. The plots of the affine 
OP strain $e_3^A$ and the affine NOP $e_1^P$ are shown in Fig(4)(b). We 
use the obtained strains to evaluate the displacements at the boundary 
and hence the shape of the system which is plotted in Fig(3)(b)(ii). 

The system is then transformed by increasing $a_3$ and 
the resulting $e_3^A$ and $e_1^A$ are plotted in Fig(4)(c). There are remnant 
(plastic) shear strains in the system even after the solid is fully transformed 
so that the system fails to go back to the initial square shape as can be seen 
from the plot of Fig(3)(c)(iii) although the crystal structure is $p4mm$ 
everywhere as expected. The shape of the solid 
is now not only determined by the affine OP strain $e_3^A$ but also, by the 
plastic part $e_3^P$ which is not recovered on increasing $a_3$. Ultimately 
this contribution causes shape irreversibility.   

\section{Summary and conclusion}
Using atomistic computer simulations of a model solid, we had shown in 
Ref.\cite{nanomart} that group-nonsubgroup transformations 
are irreversible because plasticity is automatically generated in the 
OP sector during such transformations in agreement with the conclusions of 
Bhattacharya {\em et al.}\cite{kaushik}. We showed further that if OP 
parameter plasticity is suppressed by increasing the yeild stress or in a small
system, even group-nonsubgroup transformations may become reversible. 
In this paper, we complete this line of reasoning by showing explicitly
that for shape recovery it is {\em sufficient} that all plastic strains be 
associated exclusively with the NOP sector of the transformation. If, on the 
other hand, OP plasticity is present, complete shape recovery is impossible. 
We show this here using an elastoplastic theory where it is possible to 
control the detailed nature and extent of the plastic strains by varying 
parameters. If OP plasticity is therefore ``artificially'' introduced in a 
group - subgroup ($p4mm \rightleftharpoons p2$) the normally reversible 
transformation becomes irreversible.   
Our results may be verified experimentally using shape memory alloys with 
known and non-trivial yeild criteria such that plasticity in either 
sector may be controlled independently of each other. We await such systematic
studies on the effect of plasticity on shape reversibility.  

Before we end, we would like to discuss several possible extensions of this 
work which we take up one by one as follows. 

{\em Stress induced martensite:} Within our formalism it is straightforward
to consider the effect of plastic deformation on stress induced martensitic
transformations\cite{paper0}. Stress- strain hysteresis, which can be routinely
measured in such systems may show nontrivial effects due to plasticity at 
various spatial and temporal scales. The separate effects of NOP and OP 
plasticity should be explicit in such studies which are planned in the 
near future.

{\em Scale dependent reversibility:} Reversibility of a transformation is, 
obviously, scale dependent. We have 
studied here is the question of shape reversibility, namely, reversibility
at the largest scale available to the system. In contrast, complete 
microscopic reversibility would imply that the very positions of atoms 
be recovered as the transformation is reversed - an impossibility 
considering the identity of atoms and thermal noise. At intermediate 
scales, one may ask whether microstructural features are reversible or 
not below a certain ``irreversibility length''. Such calculations are 
in progress and will be published elsewhere. 

{\em Complex dynamics and defect reorganization:} The non uniform plastic 
strain fields created during the 
transformation are redistributed as the solid is cycled through the 
transformation protocol many times. It is legitimate to ask whether this 
distribution has a steady state and what its dynamic properties may be. Defect 
redistribution and aging during transformation with nontrivial dynamical 
signatures have been observed in real systems\cite{planes}. We plan to study 
such questions, as well, within our elastoplastic approach in the future. 

\section{Acknowledgements}
Stimulating discussions with A. Saxena, K. Bhattacharya, T. Lookman, 
R. Ahluwalia, E. Salje and A. E. Jacobs are gratefully acknowledged.

\end{document}